%% file: paper.tex
\gdef\islinuxolivetti{F}
\input newhead

\input layout

\font\titlefont=ptmb at 14pt
\def\EM{E_{\rm M}}
\def\DD{{\script D}}
\def\natural{{\bf N}}
\def\a{{\bf a}}
\def\b{{\bf b}}
\def\cc{{\bf c}}
\def\d{{\rm d}}
\def\dd{{\bf d}}
\let\epsilon=\varepsilon
\def\SS{{\bf S}}
\def\ss{{\bf s}}
\parskip0pt
\parindent20pt\let\BSECTION=\SECTION
\let\BSECTIONNONR=\SECTIONNONR
\def\SECTIONNONR#1\par{\BSECTIONNONR #1\par\mark{}\write12{\folio SECTIONNONR {\noexpand#1}}}
\def\SECTION #1#2\par{\BSECTION #1\par\mark{#2}\def\rhead{\firstmark}\def\lhead{\firstmark}}

\def\rhead{}
\def\lhead{}
\def\leftheadline{{\pagenumberfont\folio}\hfil\toplinefont\lhead\hfil}
\def\rightheadline{\hfil\toplinefont\rhead\hfil{\pagenumberfont\folio}}
\headline={\ifnum\pageno>1\ifodd\pageno\rightheadline\else\leftheadline\fi\fi}
\font\tensc=ptmbc8t at 10 pt
\def\toplinefont{\tensc}
\def\pagenumberfont{\tenbf}
\setitemindent{iii)}
\null\rm
\vskip 6.3cm
\centerline{\titlefont Oscillations of Observables in}
\vskip 0.3cm
\centerline{\titlefont 1-Dimensional Lattice Systems}
\vskip 1cm
\centerline{\twelvepoint Pierre Collet\footnote{$^{1}$}
{Centre de Physique Th\'eorique, Laboratoire CNRS UPR 14,
Ecole Polytechnique, F-91128 Palaiseau Cedex (France).}
and Jean-Pierre Eckmann\footnote{$^{2}$}
{Dept.~de Physique Th\'eorique et Section de 
Math\'ematiques, Universit\'e de Gen\`eve, 
CH-1211 Gen\`eve 4 (Suisse).}}
\vskip 1cm
{\eightpoint
\noindent{\bf Abstract}: Using, and extending, striking inequalities
by V.V. Ivanov on the down-crossings of monotone functions and ergodic
sums, we give
universal bounds on the probability of finding oscillations of
observables in 1-dimensional lattice gases in infinite
volume. In particular, we study the
finite volume average of the occupation number as one runs through an
increasing sequence of boxes of size $2n$ centered at the origin.
We show that the probability to see $k$
oscillations of this average between
two values $\beta $ and $0<\alpha <\beta $
is
bounded by $C R^k$, with $R<1$, where the constants
$C$ and $R$ do {\em not} depend on any detail of the
model, nor on the state one observes, but only on the ratio $\alpha
/\beta $.}
\SECTION{Introduction}{Introduction}

In two recent 
papers, V.V. Ivanov [I1, I2] derived a novel theorem on down-crossings
of monotone functions. Theorems of this kind are useful as key
elements of ``constructive'' proofs of the Birkhoff Ergodic Theorem
[B1, B2]. For example, let $h$ be a {\em non-negative} measurable
function on $\Omega$, and
let $T$ be a measurable map $T:\Omega\to\Omega$ which preserves a probability
measure $\mu$. We denote by $s_n(\omega ) $ the sum
$$
s_n(\omega )\,=\,\sum_{j=0}^{n-1} h(T^j \omega )~.
$$
Let $\beta >\alpha >0$ be given. A down-crossing is defined as a pair of
integers $n<m$ such that
$$
s_n(\omega )/n \,\ge\,\beta ~,{\rm ~~ and ~~ } s_m(\omega
)/m \,\le\,\alpha ~.
$$
Let $\Omega _k$ denote the set of $\omega $ for which $\{s_n(\omega
)/n\}_{n=1,2,\dots} $ makes at least $k$ successive down-crossings, {\it
i.e.,} there is a sequence $n_1<m_1<n_2<m_2<\dots <n_k<m_k$, such that
each pair $n_i$, $m_i$ defines a down-crossing.
The surprising result of Ivanov is the
\CLAIM Theorem(Ivan) One has the bound
$$
\mu(\Omega _k)\,\le\, (\alpha / \beta )^k~.
\EQ(ab)
$$

Note that there is no constant in front of $(\alpha /\beta )^k$, and that
the result is independent of  $\Omega  $, $\mu$, $T$ and $h\ge 0$.
Several (relatively straightforward) generalizations and consequences
have been pointed out in [I1, I2] and in the
review paper [K]. We list some of them for the convenience of the reader.

\item{1)}If $h\in L^\infty $---there is no assumption on $h\ge0$ here
and in 2), 3) below---then, for all $\beta $ and $\alpha
=\beta -\epsilon $ one has the bound $\mu(\Omega_k)\,\le\, A e^{-Bk}$,
where $A$ and $B$ depend only on $q=\epsilon /\|h\|_\infty $. One has
$B=\OO( q^2)$.
\item{2)}If $h\in L^1$, then the bound becomes
$\mu(\Omega_k)\le C(\log k)^{1/2} / k^{1/2} $, with $C$ a function
of $\epsilon /\|h\|_1$ (when $k$ is large). This is quite similar to the older estimates
$\mu(\Omega_k) \le D\|h\|_1/ (k^{1/3} \epsilon )$, see [K].
\item{3)}The above results can easily be used to actually prove the
ergodic theorem.

\medskip
\noindent In this paper, we give a partially new proof of Ivanov's theorem, and
we extend it in such a way that it applies to 1-dimensional models of
statistical mechanics.
Indeed, it suffices to consider any translation invariant
state of a spin system [R].
To be specific, we might consider an
Ising-like model with spin $0$, $1$ (in a particle interpretation)
and long-range interaction. Then $\Omega
= \{0,1\}^\integer$, $T$ is lattice translation and $\mu $ is the
Gibbs state, not necessarily pure.
For $\omega =\{\omega_n\}_{n\in\integer}\in\Omega$,
we let $h(\omega)=\omega _0$ be the value of the spin at the site 0 and then
$s_n(\omega)/n$ has the meaning of the average ``occupation number''
on the interval $[0,n-1]$. Ivanov's theorem has then the
interpretation:
\CLAIM Proposition(ising) The probability that the mean occupation number
(as a function of the volume $n$)
makes more than $k$ oscillations between $\beta $ and $\alpha $,
$0<\alpha <\beta $ is 
bounded by $(\alpha /\beta )^k$. 

Note that this statement is independent of the spin system under
consideration, of the temperature considered, of boundary conditions
or any other parameter of the system. In particular, it also holds if
the system is not in a pure state. Thus, it is a kind of geometrical
constraint on ergodic sums, or on the fluctuations of physical
observables.
If these observables can take negative values, the results will be
modified as in 1) above, but the bound will still be exponential in
$k$.

In the statement above, we considered ``boxes'' which are given by the
intervals $[0,n-1]$. However, the statement can be extended to {\em
symmetric intervals} by the following new result:
 
Assume that $T$, as defined above, is invertible. Define
for $n\in \integer$,
$$
S_{n}(\omega )\,=\,\sum _{j=-n+1}^n h(T^j \omega )~.
\EQ(sn)
$$
We now let $\Theta_k$ denote the set of those $\omega $ for which the
sequence 
$\{S_n(\omega )/(2n+1)\}_{n=0,1,2,\dots}$ makes at least $k$
down-crossings from $\beta $ to $\alpha $, $0<\alpha <\beta $.
We will show:
\CLAIM Theorem(twosided) There are two constants $C=C(\alpha
/\beta )$ and $R=R(\alpha
/\beta )<1$ such that one has the bound
$$
\mu(\Theta_k)\,\le\, C R^k~.
$$
The constants $C$ and $R$ are independent of  $\mu$,
$\Omega $, $T$, 
and $h\ge0$.

\REMARK We will describe $R$ in Section 5, but 
note that $R<1$, $R(x)\to 0$ as $x\to 0$ and
$R(x)\approx \exp\bigl (-\OO(\epsilon/ 4^{1/\epsilon })\bigr )$ when $x
=1-\epsilon $. (This is certainly not the best possible bound.)

The \clm(twosided) can be extended to sequences of volumes which tend
to infinity in a more general way as $n\to\infty $:
Let $p_1\ge0$, $p_2\ge0$, $r_1\ge0$, $r_2\ge0$ be given integers with
$p_1+p_2>0$
and
define now
$$
S_n(\omega )\,=\,\sum_{j=-np_1-r_1}^{np_2+r_2-1} h(T^j\omega )~.
\EQ(f2)
$$
Let $\Theta_k$ be the set of $\omega $ for which the sequence
$\{S_n(\omega )/(n(p_1+p_2)+r_1+r_2)\}_{n\in\natural}$ makes at least $k$
down-crossings from 
$\beta $ to $\alpha $.
\CLAIM Theorem(fancy) There are two constants $$
C\,=\,C(p_1,p_2,r_1,r_2,\alpha
/\beta )~,\quad R\,=\,R(p_1,p_2,r_1,r_2,\alpha
/\beta )<1~,
$$ such that one has the bound
$$
\mu(\Theta_k)\,\le\, C R^k~.
$$
The constants $C$ and $R$ are independent of  $\mu$,
$\Omega $, $T$, 
and $h\ge0$.

Our paper is organized as follows. In Section 2, we show the basic
inequality,
``Ivanov's theorem'' which is used in proving \clm(Ivan) for $k=1$. 
In Section 3, we extend these results to arbitrary $k$.
In
Section 4, we use the results of Section 3 to prove \clm(Ivan) for all
$k$. 
To make the paper self-contained, we give complete proofs, even when
they are essentially just rewordings of Ivanov's work.
In
Section 5, we give the proof of \clm(twosided) and \clm(fancy).

\SECTION{A proof of Ivanov's theorem}{Ivanov's Theorem}

We consider non-decreasing (not necessarily continuous) 
functions $f$ on $\real$.
Let $E=\cup_\ell E_{\ell}$ be a closed bounded subset of $\real$ which is a
finite disjoint
union of closed intervals $E_{\ell}$.
Furthermore, we assume that $\beta $ and $\alpha $ are given constants
satisfying $\beta>\alpha>0$. 
\LIKEREMARK{Definition}Let $E'$ be a subset of $E$.
A point $x\in \real$ is said to be in the
{\em shadow} of $E'$ (relative to $E$) if it is in $E$ and if
there are two numbers $y$, $z$ in $
E'$ satisfying: 
{\parskip0pt
\item{i)}$x<y<z$,
\item{ii)}the interval $(y,z)$ is contained in $ E'$,
\item{iii)}$ f(z^-)-f(x) \le \alpha (z-x)$, and $ f(y^+)-f(x)\ge \beta
(y-x)$.
}
\REMARK This definition is slightly different from the one by Ivanov.

Let $S(E')=S(E',E)$ denote the set of $x$ which are in the shadow of
$E'$ (relative to $E$).
We assume throughout that
$E$ is a fixed set and omit mostly the second argument of $S$.
If $A$ is a set in $\real$ we let $|A|$ denote its Lebesgue measure.
The proof of \clm(Ivan) is based on the following basic bound
by Ivanov [I1,I2]:
\CLAIMNONR Ivanov's Theorem(Ivanov) Under the above hypotheses, one has
the inequality
$$
|S(E,E)|\,\le\, {\alpha \over \beta } |E|~.
\EQ(ivanov)
$$

\PROOF Our proof
relies heavily on Ivanov's
ideas, but presents some simplifications.
We will first prove the following
\CLAIM Theorem(Ivanov2) Assume $f$ is a non-decreasing,
piecewise affine, continuous function.
Then one has
the inequality
$$
|S(E,E)|\,\le\, {\alpha \over \beta } |E|~.
\EQ(ivanov2)
$$

Postponing the proof of this theorem, we now show how \clm(Ivanov2)
implies \clm(Ivanov). 
We first assume that the boundary of $E$ does not contain points of
discontinuity of $f$.
To make things clearer, we indicate the
function, and the limits of the shadow, {\it i.e.,} we write $S_{f,\alpha
,\beta }(E)$. Let $f$ be an arbitrary non-decreasing function, and let
$f_n$ be a sequence of continuous, piecewise affine, functions approximating
$f$ (pointwise). We consider the sequences
$S_{n,m}(E)\,=\,S_{f_n,\alpha  (1+1/m),\beta  (1-1/m)}(E)$, for
$n=2,3,\dots$, and large $m$.
Let $U_{p,m}=\cap_{n>p} S_{n,m}(E)$. Clearly, $U_{p,m}\subset
U_{p+1,m}$. Furthermore, every $x\in S(E)$ is in $\cap_{n>n_0(x,m)}S_{n,m}(E)$
for some $n_0(x,m)<\infty $, as one can see from the definition of shadows.
Thus, we find 
$$
S(E)\,\subset\, \cup_p U_{p,m} \,=\, \lim _{p\to\infty } U_{p,m} ~,
$$
and therefore
$$
|S(E)|\,\le\, \bigl | \cup_p U_{p,m} \bigr | \,=\,
\lim_{p\to\infty} |U_{p,m}|\,\le\, \lim_{p\to\infty } \sup_{n>p}
|S_{n,m}(E)|\,\le\, {1+1/m\over 1-1/m}\cdot{\alpha \over \beta } |E|~,
$$
by \clm(Ivanov2). Taking $m\to\infty $, the proof of \clm(Ivanov) is
complete,
when the discontinuities of $f$ do not coincide with the boundary of
$E$.

If the boundary of $E$ contains discontinuity points of $f$ we can find
for each $\ell$ a decreasing sequence of closed intervals $E^{p}_{\ell}$
such that $E\subset E^{p}_{\ell}$, $E^{p}_{\ell}$ converges to $E_\ell$ and the
boundary of each $E^{p}_{\ell}$ is made up of points of continuity of
$f$. Let $E^{p}=\cup_\ell E^{p}_{\ell}$, then obviously $E\subset
E^{p}$, hence 
$S(E)\subset S(E^p)$, and therefore
$$
|S(E)|\,\le\,\liminf_{p\to\infty} |S(E^p)|\,\le\,
 {\alpha\over \beta}\liminf_{p\to\infty} |E^p|\,=\,
{\alpha\over \beta}|E|~.
$$ 
This completes the proof of \clm(Ivanov) in all cases.

\LIKEREMARK{Proof of \clm(Ivanov2)}As we have said before, we can at
this point work with piecewise affine, non-decreasing continuous functions
defined on $\real$, with a {\em finite} number of straight pieces.

We start by defining {\em regular} and {\em maximal regular}
intervals. If $A$ is a subset of $ E$ we denote by $F(A)$ the
graph of $f$ above $A$, {\it i.e.,} $F(A)=\{(x,f(x))~|~ x\in A\}$. 

\LIKEREMARK{Definition}An interval $[a,b]$ in $\real$ is called
{\em regular} if it is contained in $ E$ and if
for all $x\in[a,b]$ one has
$$
f(a)-\beta (a-x)\,\ge\,f(x)~,~{\rm and~}
f(x)\,\ge\,f(b) -\alpha (b-x)~.
\EQ(c1)
$$

This means that the graph $F([a,b])$ lies entirely in the cone spanned
by the two straight lines of \equ(c1), see \fig(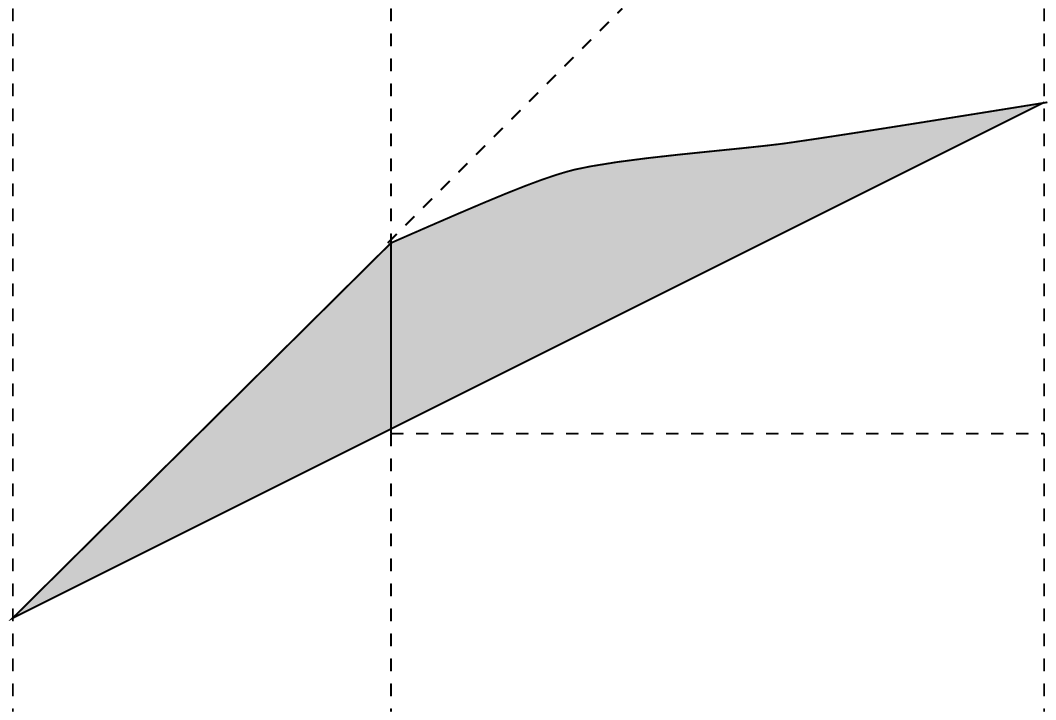).
\figurewithtexplus fig11.ps fig1.tex  8 18 -0.3 The shadow cast by a
(maximal) regular interval $[a,b]$, the cone $\CC$, and the region $\DD$.\cr
It will be useful to talk about the sets $\CC([a,b])$ and $\DD([a,b])$
spanned in this figure: Define first $c=c(a,b)$ by
$$
c(a,b)\,=\,{f(b)-f(a)+\beta a-\alpha b\over \beta -\alpha }~,
\EQ(cab)
$$
this is the $x$-coordinate of the tip of the cone.
Then we define
$$
\eqalign{
\CC([a,b])\,&=\,\{ (x,y)~|~x\in[c,a]~,~~ f(a)+\beta (x-a)\ge y \ge
f(b)+\alpha (x-b)\}~,\cr
\DD([a,b])\,&=\,\{ (x,y)~|~x\in[a,b]~,~~ f(x)\ge y\ge f(b)+\alpha
(x-b)\}~.\cr
}
$$

\LIKEREMARK{Definition}An interval $[a,b]$ in $\real$ is called
{\em maximal regular} if it is regular and is contained in no larger regular
interval. It should be noted that this definition depends on the
function $f$ {\em and} on the set $E$.

\CLAIM Lemma(disjoint) Different maximal regular intervals are disjoint.

\PROOF Since parallel lines do not intersect, one verifies easily that
the union of two regular intervals with non-empty intersection is
regular. The assertion follows.

We denote by $\EM\subset E$ the disjoint union of the maximal regular
intervals: 
$$
\EM\,=\,\cup_j \Delta_j~.
\EQ(em)
$$
The next lemma shows that it suffices to 
consider only shadows which are cast by maximal regular intervals:
\CLAIM Lemma(shadow) One has the identity $S(\EM)=S(E)$, more
precisely
$S(\EM,E)=S(E,E)$.

\PROOF If $x\in S(E,E)$, then there is at least one interval $I\subset
E$ for which $x\in S(I,E)$. By the continuity of $f$, there is a minimal
such interval in $I$, which we call $J$. This interval is regular.
The assertion follows,
because every regular interval is contained in a maximal regular
interval, as follows from the proof of \clm(disjoint).

\CLAIM Lemma(finite) The set $\EM$ is a finite union of maximal
regular intervals.

\PROOF It is here that we use the restricted class of piecewise
affine, continuous
functions. A minutes' reflection shows that the endpoints of the
$\Delta_j$ are either points of discontinuity in the slope of $f$ or
boundary points of $E$. The assertion follows because there are a
finite number of such points.

We define an auxiliary function $g$.\footnote{${}^*$}{This
definition is similar to, but different from, the one given for the
function $H$ in [I2]. Our definition makes the proofs
somewhat easier.}
For $\Delta_{j}=[a_{j},b_{j}]$, let
$
c_{j}=c(a_j,b_j)
$ as above and
define intervals $G_j(x)$ by
$$
G_j(x)\,=\,\cases{
\emptyset,&when $x\le c_j$,\cr
[f(b_j)+\alpha (x-b_j),f(a_j)+\beta (x-a_j)],& when $x\in(c_j,a_j]$,\cr
[f(b_j)+\alpha (x-b_j),f(x)],&when $x\in(a_j,b_j]$,\cr
\emptyset, & when $x>b_j$.\cr
}
$$ 
Note that $G_j(x)$ is simply
the intersection of a vertical line at $x$ with the cone
$\CC([a_j,b_j])$ or the set $\DD([a_j,b_j])$, and $|G_j(x)|$ is continuous.
We define
$$
g(x)\,=\,\bigl |\cup_j G_j(x)\bigr |~,
$$
and note that this is finite, since each $|G_j(x)|$ is bounded by
$\alpha (b_j-a_j)$, so that $g(x)/\alpha $ is bounded by the diameter
of $E$.
By construction, $g$ measures the length of the vertical cuts across the system
of cones $\CC$ and sets $\DD$
generated by the $\Delta_j$, {\em not} including multiplicities
if the cones overlap.

Our next operation consists in partitioning the shadow into those 
pieces $\Delta'_j$
generated by a $\Delta_j$ {\em under itself},
and those cast by a cone associated with a $\Delta_i$ to the right of
$\Delta _j$. In formulas:
$$
\Delta_{j}'\,=\,S(\Delta_{j})\cap \Delta_{j}\,=\,S(\Delta_{j},E)\cap \Delta_{j}~,
$$
and
$$
\Delta_{j}''=\big(S(E)\cap\Delta_{j}\big)\backslash \Delta_{j}'~.
$$
See Fig.~3 below for a typical arrangement.
We first argue that $\Delta'_j$ can be characterized by {\em
looking only at slopes $\beta $}.
\CLAIM Lemma(deltaprime) One has
$$
\Delta'_{j}\,=\,\bigl \{x\in\Delta_{j}~|~\exists
~y\in\Delta_{j}, y>x, {\rm~for~which~} 
f(y)-f(x)\ge\beta(y-x)\bigr \}~.
$$

\PROOF It suffices to show that the second set is included in
$\Delta_j'$.
Consider the ray $\{(z,f(x)+\alpha (z-x)~|~z>x\}$. If it
intersects $F(\Delta_j\cap [y,b_j])$ then $x\in S(\Delta_j)$. If not, then
$x\notin \Delta_j$, since $\Delta_j$ is regular. Hence $x\notin \Delta
'_j$ either and the proof is complete.

We now can use the Riesz lemma to give a {\em bound} on the size
of $\Delta_j'$:
\CLAIM Lemma(Riesz) One has the inequality
$$
|\Delta_{j}'|\,\le\, {f(b_{j})-f(a_{j})\over \beta}~.
\EQ(r1)
$$

\PROOF Define $s(x)=f(x)-\beta x$. Then, by \clm(deltaprime), we see
that 
$$
\Delta'_{j}\,=\,\bigl \{x\in\Delta_{j}~|~\exists
~y\in\Delta_{j}, y>x, {\rm~for~which~} 
s(y)\ge s(x)\bigr \}~.
$$
We apply here a variant of the
Riesz lemma [RN, Chapter 1.3].\footnote{*}{The Riesz lemma is
formulated in [RN] for arbitrary functions, with {\em open} intervals. Because
we have piecewise affine functions, we can go over the proof and
obtain the result for closed intervals.} It tells us that $\Delta'_j$ is
a finite disjoint union
$$
\Delta'_j\,=\,\cup _k [a_{j,k},b_{j,k}]~,
$$
and that furthermore, for every of these intervals one has the
inequality
$$
s(x)\,\le\,s(b_{j,k})~,
$$
when $x\in[a_{j,k},b_{j,k}]$. Taking $x= a_{j,k}$, we get
$$
f(a_{j,k})-\beta a_{j,k}\,\le\,f(b_{j,k})-\beta b_{j,k}~,
$$
and thus
$$
|\Delta'_j|\,=\,\sum_k (b_{j,k}-a_{j,k})\,\le\,\beta ^{-1} \sum_k \bigl
(f(b_{j,k})-f(a_{j,k})\bigr )\,\le\,\beta ^{-1}\bigl (
f(b_j)-f(a_j)\bigr )~.
$$
The last inequality is a consequence of the monotonicity of $f$.
The proof of \clm(Riesz) is complete.

We next study $\Delta''_j$.
\CLAIM Lemma(deltapp) One has the following inequality:
$$
\beta |\Delta_{j}''|\,\le\,
 {g(b_{j})-g(a_{j})}
-f(b_{j})+f(a_{j})+\alpha(b_{j}-a_{j})~.
$$

\PROOF First observe that if $x\in\Delta''_{j}$, then by
\clm(deltaprime)
the infinite ray
$$
\{(x+s,f(x)+\beta s)~|~ 0<s\}
\EQ(ray)
$$ 
does {\em not} meet the graph $F(\Delta_{j})$. Consider next any
vertical line. To be specific, we take the line whose abscissa is
$b_j$,
and, since each of the previous rays
emanates from a unique point of $F(\Delta_j)$,
this provides a bijection between $\Delta''_{j}$ 
and its projection $D''_j$ along the
slope $\beta$ onto the vertical line of abscissa $b_{j}$. See
\fig(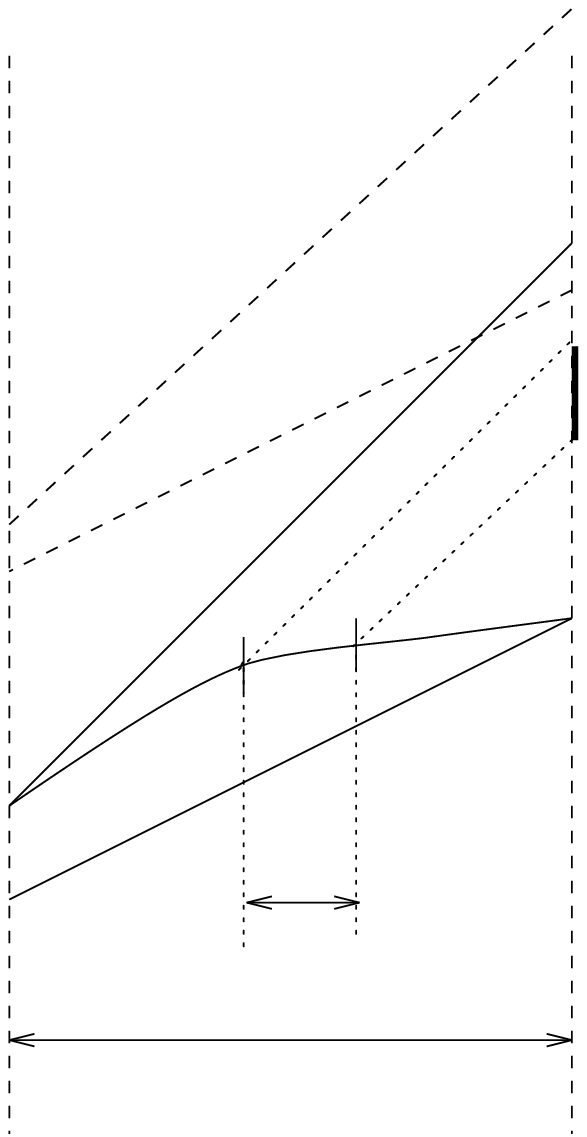).
\figurewithtexplus fig21.ps fig2.tex 11.5  8.5 0.6 The bijection between
$\Delta_j''$ and $D_j''$. The size of $\Delta''_j$ is taken here
symbolically.
See \fig(fig31.ps) for a realistic arrangement.\cr
Note that $D''_j$ is a union of disjoint intervals and satisfies
$|D''_j|=\beta |\Delta''_j|$. To understand the following
construction, it is useful to consider \fig(fig31.ps).
\figurewithtexplus fig31.ps fig3.tex 12.5  15.5 0.8 The region $ABCD$,
through which a cone passes. The intersection of the cone with the vertical line at
$b$ is $G(b)$. Inside this cone there is the bijection
between $\Delta''$ (which has 2 pieces)
and $D''$, and there is a piece of shadow, $\Delta'$, which
is generated from the (maximal) regular interval $[a,b]$ itself. Note
that $G(a)$ and $G(b)$ will in general contain pieces from other cones as
well.\cr
Consider a fixed
$\Delta_j$, we will omit the index $j$ in this argument. 
We define two intervals:
$$
\eqalign{
Q(a)\,&=\,[f(b)+\alpha (a-b),f(a)]~,\cr
Q(b)\,&=\,[f(b),f(a)+\beta (b-a)]~,\cr
}
$$
and we let $q(a)=|Q(a)|$. We have the following chain of inequalities:
\medskip
{\parskip0pt
\item{1)} $g(a)-q(a)\le | G(a)\setminus Q(a)|$,
\item{2)}  $| G(a)\setminus Q(a)|\le  | G(b)\setminus Q(b)|$,
\item{3)} $ | G(b)\setminus Q(b)| \le | G(b)\setminus D''|$,
\item{4)} $| G(b)\setminus D''| \le g(b) -| D''|$.
}
\medskip
\noindent Inequality 1) follows from $Q(a)\subset G(a)$, 3) follows
from $D''\subset Q(b)$ and 4) from $D''\subset G(b)$ which holds by
the definition of $\Delta''$ and the bijection constructed above.
The inequality 2) describes the intersections of the cones outside of
the interesting sets $Q(a)$ resp.~$Q(b)$. If the cones do not
intersect $ABCD$ in \fig(fig21.ps), the statement is trivial. If they
intersect this region partially, the statement follows by examining
the (rather obvious) cases which can occur.

Combining 1)--4), we see that
$$
\beta |\Delta_j''|\,=\,|D_j''|\,\le\,g(b_j) -g(a_j)+q(a_j)~.
\EQ(ineq)
$$
Since $q(a_j)=f(a_j)-f(b_j)+\alpha (b_j-a_j)$, the
claim \clm(deltapp) follows. 

Combining \clm(Riesz) and \clm(deltapp), and using again the
definition of $q(a_j)$,
we get immediately
\CLAIM Corollary(deltabound) One has the bound
$$
|S(E)\cap \Delta_{j}|\,\le\, {g(b_{j})-g(a_{j})\over \beta } +{\alpha\over\beta}
|\Delta_{j}|~. 
$$

We next consider a maximal interval $E'=[a',b']$ 
of $E\backslash
\EM$.
\CLAIM Lemma(outer) One has the inequality
$$
|S(E)\cap E'|\,\le\, {g(b')-g(a')\over \beta }+{\alpha \over \beta
}|E'|~.
\EQ(outer)
$$

\PROOF We
distinguish two cases. 
Assume first that at least one cone ``traverses'' $E'$ completely,
{\it i.e.,} its tip ``$c$'' is to the left of the interior of $E'$ and its
point ``$a$'' is to the right.
Then
$$
|E'|\,=\,b'-a'\,\le\, {g(b')-g(a')\over \beta -\alpha }~,
$$
or equivalently
$$
\beta |E'|\,\le\, \alpha |E'|+{g(b')-g(a') }~.
$$
Since $S(E)\cap E'\subset E'$ the assertion follows.
If no cone traverses $E'$ completely, but some penetrate into it,
we consider instead of the interval $S(E)\cap E'$ the
shortest subinterval $[c,b']$ containing the projection of all the
cones
onto the $x$-axis. Since $S(E) \cap E' \subset [c,b']$, the
assertion follows as before.

It is now straightforward to complete the proof of \clm(Ivanov2):
First observe that if $X=[x_1,x_2]$ is an interval of $\real\setminus
E$, then
$0=|S(E,E)\cap X|\le g(x_2)-g(x_1)$, since the widths of the cones is
increasing in the gaps of $E$.
Combining this with \clm(deltabound) and \clm(outer), and observing that the
intervals $E'$, $\Delta'_j$, and $X$ have contiguous boundaries, we get
a telescopic sum in which the $g(\cdot)$ all cancel, except the first
and the last. The first is subtracted, and the last is zero. The other
terms add up to $(\alpha /\beta )|E|$, and the proof is complete.

\SECTION{The iterated theorem}{The Iterated Theorem}

We now give a bound, analogous to \clm(Ivanov) for the case of $k$
oscillations.
\CLAIM Theorem(kIvanov) Let $E_k$ the set of $x\in E$ for which the
function
$f$ has $k$ successive
down-crossings---as defined in Section 1---from $\beta $ to $\alpha <\beta $ to the
right of $x$. Then
$$
|E_k|\,\le\,(\alpha /\beta )^k |E|~.
$$

\PROOF The case $k=1$ is an immediate consequence of \clm(Ivanov),
because if $x$ is in $S(E)$ it is in the shadow of some regular
interval $J$, and this means there is (at least) one down-crossing from
$\beta $ to $\alpha $.
The proof proceeds by induction. Assume we have shown the claim for all
$k<k^*$. If $x\in E_{k^*}$, we let
$[y_j,z_j]$, $j=1,\dots,k^*$ denote the intervals of successive crossings.
Each of the cones $\CC([y_j,z_j])$ contains a smaller cone which has its apex at the point
$(x,f(x))$.
Therefore $x$ is in the shadow of all the other
cones. But this means that if $x\in E_{k^*}$ then $x\in S(E_{k^*-1})$. The
assertion follows.
\SECTION{Proof of \clm(Ivan)}{Proof of \clm(Ivan)}

We first need to define
the notion of down-crossing of sequences more precisely.
\LIKEREMARK{Definition}For every $\beta >\alpha >0$ and every $k\in\natural$
we define $C_{k,\alpha ,\beta }$ as the set of monotone sequences
$\cc=\{c_n\}_{n=0,1,\dots}$ for which $\{c_n/n\}_{n\in\natural}$ makes
$k$ down-crossings from $\beta $ to $\alpha $:
$$
\eqalign{
C_{k,\alpha ,\beta }\,&=\,
\biggl\{ \{c_n\}_{n\ge0}~|~
c_j\ge c_{j-1} {\rm ~for~}j=1,2,\dots~,\cr
&{\rm there~are~numbers~~}0<n_1<m_1<n_2<m_2<\cdots< m_k{\rm ~~for~which}\cr
& c_{n_i}/n_i\ge \beta ,~~c_{m_i}/m_i\le \alpha,{\rm
~for~}i=1,\dots,k
\biggr\}~.\cr
}
\EQ(defc)
$$
We shall say that $\cc\in C_{k,\alpha ,\beta }$ has $k$ oscillations of
amplitude $\beta /\alpha $.\footnote{${}^*$}{This terminology is adequate
since all bounds will be functions of the amplitude $\beta /\alpha $
alone, {\it i.e.,} they only depend on the relative size of
$\alpha $ and $\beta $.}

Given a sequence $\cc$, and $\ell\ge0$, we define a new sequence
$\dd^{(\ell,L)}$ by $d_n^{(\ell,L)}=c_{n+\ell}-c_{\ell}$, $n=0,\dots,L-\ell$.
We denote by $I(\cc,k,\alpha ,\beta ,L)$ the set of those indices
$\ell$, for which $\dd^{(\ell,L)} \in C_{k,\alpha ,\beta }$.
Thus, $I(\cc,k,\alpha ,\beta ,L)$ counts how many ``shifted''
subsequences of $\{c_0,\dots,c_L\}$ make at least $k$ oscillations.
In other words, for $\ell\in I(\cc,k,\alpha ,\beta ,L)$, the sequence
$$
\left \{ c_{n+\ell}-c_\ell\over n\right \}_{n=0,\dots,L-\ell}~,
$$
makes at least $k$ down-crossings between $\beta $ and $\alpha $.
\CLAIM Proposition(seq1) One has the inequality:
$$
|I(\cc,k,\alpha ,\beta ,L)|\,\le\,(\alpha /\beta )^k (L+1)~.
$$

\REMARK See Ivanov [I1] for the manipulations---essentially a
``periodic'' extension of 
the sequence $\{c_0,\dots c_L\}$---which lead to the bound
$(\alpha /\beta )^k L$.
\PROOF We apply \clm(kIvanov) to the following setting. We let
$E=[0,L+1)$, and we let $f(x)=c_j$ for $x\in [j,j+1)$.
It is easy to verify that if an index $j$ is such that the sequence
$\cc$ has $k$ down-crossings from $\beta$ to $\alpha$ to the right of
$j$, 
then the same is true for the function $f$ on the interval $[j,j+1)$. In
other words,
$$
|I({\bf c},k,\alpha,\beta,L)|\,\le\, |E_k|~,
$$
and the result follows from Theorem 3.1.

\LIKEREMARK{Proof of \clm(Ivan)}At this point, we use the invariance
of the measure $\mu$ under $T$. 
For every $\omega \in\Omega $,
we consider sequences $\ss(\omega )=\{s_n(\omega )\}$, where
$s_n(\omega )= \sum_{\ell=0}^{n-1}f(T^\ell \omega )$. We let 
$
\Omega_{k,\alpha ,\beta }
$
denote the set of those $\omega $ for which the sequence $\ss(\omega )$
makes $k$ oscillations of amplitude $\beta /\alpha $, and we let
$\Omega _{k,\alpha ,\beta ,m}$ be the subset of those $\omega $ where
this happens for the subsequence $\{s_1(\omega ),\dots,s_m(\omega )\}$.
We then have, since $\mu(A)=\mu(T^{-1}A)$,
$$
\eqalign{
X\,\equiv\,\mu(\Omega _{k,\alpha ,\beta })\,&=\,
\lim_{m\to\infty } \mu(\Omega _{k,\alpha ,\beta ,m})\cr
\,&=\,\lim_{m\to\infty } L^{-1}
\sum_{j=0}^{L-1} \mu ( T^{-j}\Omega _{k,\alpha ,\beta ,m})\cr
\,&=\,\lim_{m\to\infty } L^{-1}
\int \d\mu(\omega )\sum_{j=0}^{L-1} \chi
_ {T^{-j}\Omega _{k,\alpha ,\beta ,m}}(\omega )\cr
\,&=\,\lim_{m\to\infty } L^{-1}\int \d\mu(\omega )\sum_{j=0}^{L-1} \chi
_ {\Omega _{k,\alpha ,\beta ,m}}(T^j\omega )\,\equiv\,\lim_{m\to\infty
}X_{m,L}~.
}
\EQ(ss1)
$$
Note now that
$
 \chi
_ {\Omega _{k,\alpha ,\beta ,m}}(\omega' )=
1$,
if the sequence $\{s_n(\omega ')\}_{n=1,\dots,m} $
makes $k$ oscillations of amplitude $\beta /\alpha $, and 0
otherwise.

The crucial observation by Ivanov is now that if
$$
\chi
_ {\Omega _{k,\alpha ,\beta ,m}}(T^j\omega )=1,{\rm~ then~~}
j\in I(\ss(\omega ),k,\alpha ,\beta ,L+m-1)~,
\EQ(trick)
$$ as one can see just from
the definitions.
Therefore, by \clm(seq1), we find
$$
\sum_{j=0}^{L-1} \chi
_ {\Omega _{k,\alpha ,\beta ,m}}(T^j\omega )\,\le\,
| I(\cc(\omega ),k,\alpha ,\beta ,L+m-1)|\,\le\, (\alpha /\beta )^k
(L+m)~.
$$
Coming back to $X_{m,L}$, we see that
$$
\eqalign{
X_{m,L}\,&\le\,L^{-1}\int \d\mu(\omega ) (\alpha /\beta
)^k \cdot (L+m)~,\cr
}
$$
{\em for all $L$}, and therefore
$$
\eqalign{
X_m\,\equiv\,\limsup_{L\to\infty }X_{m,L}\,&\le\,\limsup_{L\to\infty }~(\alpha /\beta )^k
{L+m\over L}
\,=\,(\alpha /\beta )^k~.\cr}
\EQ(ss2)
$$
Since $X\le \lim_{m\to\infty } X_m$,
the assertion of \clm(Ivan) follows.

\SECTION{Symmetric intervals}{Symmetric Intervals}

In this section, we prove \clm(twosided), and \clm(fancy).
The proofs leading to
\clm(Ivan) are not quite applicable, because the device used in
Eq.\equ(trick) does not work in the case of symmetric intervals,
since a subsequence will cut a ``hole'' in the original sequence.
However, we shall work with the decomposition of the sequence
$S_n(\omega )=\sum_{j=-n}^{n-1} h(T^j\omega )$ as the sum of two
sequences $\a$ and $\b$ to be defined below. We first show that if
$s_n$ oscillates, then at least one of the sequences $\a$ or $\b$ must
oscillate as well, but a little less. We study this as a general problem:

We assume $\cc=\{c_n\}_{n\ge0}\in C_{k,2\alpha ,2\beta }$ and
further that $c_n=a_n+b_n$, where $\a=\{a_n\}$ and $\b=\{b_n\}$
are monotone sequences of non-negative numbers. We are going to show
that {\em either $\a$ or $\b$ must have oscillations}, and we will
give bounds on the number and size of these oscillations. (Our bounds
are not optimal, and we do not know the optimal bounds, but we will
give a reasonable set of bounds for the cases when $\alpha /\beta $
is close to 0 or 1.)

To describe the nature of the oscillations, we set
$$
\tau \,=\,{1+(\beta/\alpha)\over 2} ~,
$$
so that $1<\tau <\beta /\alpha $.
Then we define for $j=1,2,\dots$,
$$
\eqalign{
\alpha _j\,&=\,\alpha +2(j-1)(\alpha -\beta /\tau)~,\cr
\beta _j\,&=\,\tau\alpha_j~,\cr
\gamma_j\,&=\,2\beta /\tau -\alpha -2(j-1)(\alpha -\beta /\tau )~.
}
\EQ(abdef)
$$
We also define $k_0=k$ and $k_n=1+[ k_{n-1}/2^n]$, where $[\ ]$
denotes the integer part.
We can now formulate our result:
\CLAIM Proposition(oscill) If $\cc\in C_{k,2\alpha ,2\beta }$ and
$\cc=\a+\b$ as above, then at least one of the sequences $\a$ or $\b$
is in
$$
C'_{k,\alpha ,\beta }\,\equiv\,
\left (\bigcup _{p^*\ge n\ge 1} 
C_{k_{2n+1},\gamma _n,\tau \gamma _n}\right ) ~\bigcup~
\left (\bigcup _{p^*\ge n\ge1} C_{k_{2n},\alpha _n ,\tau \alpha _n}\right )~,
$$
where $p^*$ is the smallest integer satisfying
$$
p^*\,\ge\,{\alpha +\beta  \over 2(\beta -\alpha )}+1~.
$$

\REMARK The meaning of this inclusion is that either $\a$ or $\b$ make
at least $k_{2p^*+1}$ oscillations of ``amplitude'' $\tau $.
Thus, the theorem says that if $\cc$ has
$k$ oscillations of amplitude $\beta /\alpha $, then, for large $k$,
{\em $\a$ or $\b$ have at least $\OO(k/4^{p^*})$ oscillations of
amplitude $\tau $}. Note that if $\beta /\alpha $ diverges then $\tau $
diverges as well, while for $\beta /\alpha =1+\epsilon $ we have $\tau
=1+\epsilon /2$.

\PROOF Before we start with the proof, we note that the definitions of
$\alpha _j$, $\beta _j$ have been chosen such that for $j\ge1$, one
has
$$
\tau \alpha _j\,=\,\beta _j~,
\quad
\tau \gamma_j\,=\,2\beta-\beta_j~, \quad \alpha _{j+1}\,=\,2\alpha
-\gamma _j~.
\EQ(why)
$$
We will construct recursively the possible sets of indices for
which oscillations occur. Assume $\cc\in C_{k,2\alpha ,2\beta }$,
with the oscillating indices $m_j$, $n_j$ as in Eq.\equ(defc). Define
$I_0=J_0=\{1,\dots,k\}$,
and
$$
\eqalign{
J_0^a\,&=\,\{i\in J_0~|~ a_{m_i} \le \alpha  _1 m_i \}~,\cr
J_0^b\,&=\,\{i\in J_0~|~ b_{m_i} \le \alpha _1 m_i \}~.\cr
}
$$
Since $a_{m_i}+b_{m_i}=c_{m_i}\le 2 \alpha m_i = 2\alpha_ 1 m_i$, we
see that each $i\in J_0$ must be in at least one of the sets $J_0^a$,
$J_0^b$. Therefore the cardinalities satisfy $|J_0^a|+|J_0^b|\ge
|J_0|=k=k_0$, and we conclude that $\max(|J_0^a|,|J_0^b|) \ge k_1$. We
assume for definiteness that $|J_0^a|\ge k_1$; in the other case, the
proof is obtained by exchanging the r\^oles of $\a$ and $\b$. 
We define next
$$
\eqalign{
I_1^a\,&=\,\{i\in J_0^a~|~ a_{n_i} \ge \beta   _1 n_i \}~.\cr
}
$$
Assume first
$|I_1^a|\ge k_2$. By the definition of $J_0^a$ and $I_1^a$,
this means---cf.\ Eq.\equ(why)---that $\a\in C_{k_2,\alpha _1,\beta
_1}=C_{k_2,\alpha _1,\alpha _1\tau }$,
which is part of the set $C'_{k,\alpha ,\beta }$, and we stop
the induction. In the other case, we 
define
$I_1^b=J_0^a\setminus I_1^a$. Clearly, $|I_1^b|\ge k_2$, but
furthermore we have for all $i\in I_1^b$ the inequalities
$$
\eqalign{
a_{n_i}\,&< \,\beta _1 n_i~,\cr
a_{n_i}+b_{n_i}\,&\ge\, 2\beta n_i~,\cr
}
$$
and therefore
$$
b_{n_i}\,\ge\,(2\beta-\beta _1) n_i~.
\EQ(x1)
$$
We now define
$$
J_1^b\,=\,\{i\in I_1^b~|~ b_{m_i}\le\gamma _1 m_i\}~.
$$
If $|J_1^b|\ge k_3$, then we have, using Eqs.\equ(x1) and \equ(why),
$$
\b\,\in\, C_{k_3, \gamma _1, 2\beta-\beta _1}\,=\,
C_{k_3,\gamma_1,\tau\gamma _1}~,
$$
and we stop the induction.
In the other case, we let $J_1^a=I_1^b\setminus J_1^b$, and then for
all $i\in J_1^a$ we have 
$$
\eqalign{
b_{m_i}\,&>\,\gamma_1 m_i~,\cr
a_{m_i}+b_{m_i}\,&\le\,2\alpha m_i~,\cr
}
$$
and therefore
$$
a_{m_i}\,\le\,(2\alpha-\gamma_1) m_i\,=\,\alpha _2 m_i~.
\EQ(maybe)
$$
If $2\alpha -\gamma_1<0$, the inequality \equ(maybe) contradicts the
positivity of the $a_j$ and hence $|J_1^b|< k_3$ will never occur
and the induction stops.

Otherwise, we continue, defining for $\ell\ge 2$,
$$
\eqalign{
I_\ell^a\,&=\,\{ i\in J_{\ell-1}^a ~|~ a_{n_i}\ge \beta _{\ell } n_i\}~,\cr
I_\ell^b\,&=\,J_{\ell-1}^a\setminus I_\ell^a~,\cr
J_\ell^b\,&=\,\{ i\in I_{\ell}^b ~|~ b_{m_i}\le \gamma_{\ell } m_i\}~,\cr
J_\ell^a\,&=\,I_{\ell}^b\setminus J_\ell^b~.\cr
}
$$
There are now four cases. 
\item{1)}If $|I_\ell^a|\ge k_{2\ell}$, then
$I_\ell^a\subset J_{\ell-1}^a$ implies $a _{n_i}\ge\beta _{\ell } n_i$
and $a_{m_i}\le \alpha _{\ell } m_i$ for $i\in
I_\ell^a$, and hence $\a\in C_{k_{2\ell},\alpha
_{\ell },\beta_{\ell }}= C_{k_{2\ell},\alpha_{\ell },\tau \alpha 
_{\ell }} $, and the induction stops.
\item{2)}If $|I_\ell^a|<k_{2\ell}$, then we have
for $i\in I_\ell^b$ the inequality $b_{n_i}\ge (2 \beta - \beta _{\ell })n_i$,
since $a_{n_i}<\beta _{\ell } n_i$ and $a_{n_i}+b_{n_i}\ge 2\beta n_i$,
and we continue the induction.
\item{3)}If $|J_\ell^b|\ge k_{2\ell+1}$, then
$J_\ell^b\subset I_{\ell}^b$ implies $b _{m_i}\le\gamma_{\ell } m_i$
and $b_{n_i}\ge (2\beta-\beta _{\ell }) n_i$ for $i\in
J_\ell^b$, and hence $\b\in C_{k_{2\ell+1},\gamma
_{\ell },2\beta-\beta_{\ell }}= C_{k_{2\ell+1},\gamma_{\ell },\tau\gamma
_{\ell } } $, and the induction stops.
\item{4)}In the last case, $|J_\ell^b|<k_{2\ell+1}$, and then we have
for $i\in J_\ell^a$ the inequality $a_{m_i}\le (2 \alpha -
\gamma_{\ell })m_i $,
since $b_{m_i}>\gamma_{\ell }m_i$ and $a_{m_i}+b_{m_i}\le 2\alpha
m_i$.
If $ (2 \alpha -
\gamma_{\ell })\ge0$, we continue the induction, while in the
opposite case, we see that  $|J_\ell^b|< k_{2\ell+1}$ cannot
occur, and the induction stops.

\medskip\noindent
Since $2\alpha -\gamma_{p^*}<0$, as one checks easily from the
definitions, the induction must stop for some $\ell\le p^*$.
The proof of \clm(oscill) is complete.

We can now complete the proof of \clm(twosided) by applying
 \clm(oscill).
We write the sum
$S_n$ of Eq.\equ(sn) as
$$
S_n(\omega )\,=\, a_n(\omega )+b_n(\omega )~,
$$
where
$$
a_n(\omega )\,=\,\sum_{j=0}^{n-1} h(T^j\omega )~,
\quad
b_n(\omega )\,=\,\sum_{j=1}^{n} h(T^{-j}\omega )~.
$$
By \clm(oscill), if $\SS(\omega )\in C_{k,2\alpha ,2\beta }$ then 
at least one of the sequences $\a(\omega )$, $\b(\omega )$ is in $C'_{k,\alpha
,\beta }$.
Therefore
$$
\mu(\{\omega ~|~ {\bf S}(\omega )\in C_{k,\alpha ,\beta }\})\,\le\,
\mu(\{\omega ~|~ {\bf a}(\omega )\in C'_{k,\alpha ,\beta }\})\,+
\mu(\{\omega ~|~ {\bf b}(\omega )\in C'_{k,\alpha ,\beta }\})
~.
$$
Since $\mu$ is invariant under $T$ and $T^{-1}$, we can apply
\clm(Ivan) to {\em both} sequences and we get a bound:
$$
\mu(\{\omega ~|~ {\bf S}(\omega )\in C_{k,\alpha ,\beta }\})\,\le\,
2\sum _{n=1}^{2p^*+1}  (1/\tau) ^{k_n}\,\le\, 4(p^* +1)(1/\tau
)^{k/4^{p^*+1}}~.
$$
Since both $\tau $ and $p^*$ are functions of $\alpha /\beta $ and
$\tau >1$, the \clm(twosided) follows.

\LIKEREMARK{Proof of \clm(fancy)}This proof will be
straightforward combination of the 2 following lemmas.
\CLAIM Lemma(s1) Let $p\ge0$. There are a $k'=k'(p,\alpha /\beta )$
and a $\beta   '=\alpha   B(\alpha /\beta )$, with $B>1$ when $\alpha
<\beta $, such that if $\{q_n\}\in
C_{k,\alpha ,\beta }$, then the sequence with elements
$t_n=q_{n}{n\over n+p}$ is in
$C_{k-k',\alpha ,\beta '}$.

\REMARK It will be obvious from the proof that similar statements hold
in the following cases:
$$
\eqalign{
\{t_n\}\,&=\, \{q_n \max(0,(n-p))/n\}  \in C_{k-k',\alpha ,\beta
'}~,\cr
\{t_n\}\,&=\, \{q_n n/ \max(1,(n-p))\}  \in C_{k-k',\alpha' ,\beta
}~,\cr
\{t_n\}\,&=\, \{q_n (n+p)/n\}  \in C_{k-k',\alpha' ,\beta
}~,\cr}
\EQ(all)
$$
where $\alpha '= \beta A(\alpha /\beta )$ with $A<1$ if $\alpha <\beta
$.
\PROOF We will actually construct $k'$ and $\beta '$. Let $n_i$ and
$m_i$ be defined as the crossing points of the sequence $s_n$,
cf.~Eq.\equ(defc). Since $n_1\ge 1$, and the $s_n$ form an increasing
sequence, we have
$$
\alpha m_i\,\,\ge\,s_{m_i}\,\ge\,s_{n_i}\,\ge\,\beta n_i~,
$$
so that $m_i\ge (\beta /\alpha )n_i>(\beta /\alpha ) m_{i-1}$ and thus
$$
m_i\,\ge\, (\beta /\alpha )^i ~.
\EQ(mi)
$$
Therefore, 
$$
t_{n_i}\,=\,s_{n_i}{n_i\over n_i+p}\,\ge\,\beta n_i {n_i\over n_i+p}\,=\,\beta
n_i (1+{p\over n_i})^{-1}\,\ge\,
n_i {\beta \over  1+p(\alpha /\beta )^{i-1}}~.
$$
We choose 
$$
{B(\alpha /\beta ) }\,=\,{1+(\beta/\alpha) \over 2}~,
$$
so that $\beta '=\alpha B(\alpha /\beta )>\alpha $, and there is clearly a
$k'=k'(p,\alpha /\beta 
)$ for which $\beta /\bigl (1+p(\alpha /\beta )^{k'-1}\bigr )>\beta '$. Then we
have for $i>k'$,
$$
t_{n_i}\,\ge\, n_i \beta '~.
$$
On the other hand,
$$
t_{m_i}\,=\,q_{m_i}{m_i\over m_i+p}\,\le\, q_{m_i} \,\le\,\alpha m_i ~,
$$
so that the assertion follows.

We next study sequences with increments of more than 1. Fix $r\in{\bf
N}$
and define
$$
t_n(\omega )\,=\,\sum_{j=0}^{rn-1} h(T^j\omega )~.
$$
We are interested in the oscillations of $t_n/(nr)$. This question is
reduced to the one described in \clm(ising): Let
$$
h_r(\omega )\,=\,{1\over r}\sum_{j=0}^{r-1} h(T^j\omega )~,\quad
T_r\,=\,T^r~,
$$ and
$$
s_n(\omega )\,=\,\sum_{j=0}^{n-1} h_r\bigl ((T_r)^j\omega \bigr )~.
$$
By construction, $s_n(\omega )=t_n(\omega )$.
Since $h_r\ge0$ and $T_r$
preserves the measure $\mu$ if $T$ preserves it, we conclude
\CLAIM Lemma(multi) The probability that
the sequence $\{t_n/(nr)\}$ (defined with $h$ and $T$) makes
at least $k$ oscillations is the same as the probability that $\{s_n/n\}$
(defined with $h_r$ and $T_r$) makes at least $k$ oscillations, and this
quantity is bounded by $(\alpha /\beta )^k$.

\REMARK The \clm(multi) is a little too strong for our purpose, since
it would have sufficed to observe that the sequence $\{s_n/n\}$ makes
more oscillations than $\{t_n/(nr)\}$.

We can now complete the proof of \clm(fancy) by a painful but somehow
obvious combination of the results above.
Recall the definition of $S_n$ in Eq.\equ(f2):
$$
S_n(\omega )\,=\,\sum_{j=-np_1-r_1}^{np_2+r_2-1} h(T^j\omega )~.
$$
We want to bound the probability that the sequence
$S_n/\bigl(n(p_1+p_2)+r_1+r_2\bigr)$ makes $k$ down-crossings from $\beta $ to
$\alpha $. So assume the sequence with elements $q_n\equiv n\cdot S_n/\bigl(
n(p_1+p_2)+r_1+r_2\bigr)$ is in $C_{k,\alpha ,\beta }$.
We let
$$
t_n\,=\,q_n{n+p\over n}\,=\,{S_n\over p_1+p_2}~, {~~\rm where~~} p\,=\,{r_1+r_2\over p_1+p_2}~.
$$
Applying \clm(s1), (actually Eq.\equ(all)), we see that the sequence
with elements $S_n/(p_1+p_2)$ is in $ C_{k-k',\alpha ',\beta }$, and
thus the sequence with elements $S_n$ is in $C_{k'',\alpha '',\beta
''}$, where $k''=k-k'$, $\alpha ''=\alpha '/(p_1+p_2)$, $\beta''=
\beta/(p_1+p_2)$. 
We next use the ``splitting'' mechanism and write $S_n=a_n+b_n$, where
$$
a_n\,=\, \sum_{j=1}^{np_1+r_1} h(T^{-j}\omega)~, {\rm~~and~~}
b_n\,=\, \sum_{j=0}^{np_2+r_2-1} h(T^{j}\omega)~.
$$
By \clm(oscill), we conclude that one of the two sequences $\a=\{a_n\}$ or
$\b=\{b_n\}$ must oscillate; we discuss here the case where it is $\a$ and
leave the other case to the reader.
Then we conclude that there are a $k^{(3)}$, $\alpha^{(3)}$ and
$\beta^{(3)}$
for which $\a\in C_{k^{(3)}, \alpha^{(3)} ,
\beta^{(3)}}$ and these constants depend only on $\alpha /\beta$, and
furthermore $k^{(3)}=\OO(k)$ as $k\to\infty $. Finally, $\alpha^{(3)}
/\beta^{(3)}<1$ when $\alpha <\beta $. (We will construct further such
constants and they will possess the same properties. Of course, with
some more work one can see that the quotient $\alpha^{(3)}
/\beta^{(3)}$ goes to 0 when $\alpha /\beta \to 0$.)
If $\a\in C_{k^{(3)}, \alpha^{(3)} ,\beta^{(3)}}$, 
then the sequence with elements $a_n/p_1$ is in $C_{k^{(3)}, \alpha^{(3)}/p_1 ,
\beta^{(3)}/p_1}$, and, applying again Eq.\equ(all), we see that the
sequence
with elements $(a_n/p_1)\cdot n/\bigl(n+(r_1/p_1)\bigr )$ is in $C_{k^{(4)}, \alpha^{(4)} ,
\beta^{(4)}}$. This means that the sequence
with elements
$$
{s_m(\omega)\over m}\,=\,{1\over np_1 + r_1} \sum_{j=1}^{np_1+r_1} h(T^{-j}\omega)~,
$$
where $m=np_1+r_1$,
makes at least $k^{(4)}$ down-crossings from $\beta^{(4)}$ to $\alpha^{(4)}$.
The probability that this happens for $m=r_1,r_1+n,r_1+2n,\dots$ is
certainly less than the probability that this happens for the sequence
$s_m(\omega)/m$ when $m=1,2,\dots$.
But this probability is bounded, using \clm(Ivan), 
by $\bigl(\alpha^{(4)} /\beta^{(4)}\bigr)^{k^{(4)}}$. Since
$s_m(\omega)$ has been derived from the original sequence $S_n(\omega
)$ by
successive modifications, the proof of \clm(fancy) is complete.
\LIKEREMARK{Acknowledgments}Our collaboration has been made possible
through the kind hospitality of the IHES. Further support was received
from the Fonds National Suisse. This work was completed while one of
us
(JPE) was profiting from the hospitable atmosphere of Rutgers and the 
IAS in Princeton.
\vfill\eject
\LIKEREMARK{References}
\vskip0.5cm
{\eightpoint
\widestlabel{[RN]}
\ref 
\no B1
\by E. Bishop
\paper An upcrossing inequality with applications
\jour Michigan Math. J.
\vol 13
\pages 1--13
\yr 1966
\endref
\ref 
\no B2
\by E. Bishop
\paper A constructive ergodic theorem
\jour J. Math. Mech.
\vol 17
\pages 631--639
\yr 1968
\endref
\mark{}
\ref
\no I1
\by V.V. Ivanov
\paper Oscillations of means in the ergodic theorem.
(Translated from Dokl. Acad. Nauk {\rm 347}, 736--738 (1996))
\jour Russian Acad. Sci. Dokl. Math.
\vol 53
\yr 1996
\pages 263--265
\endref
\ref
\no I2 
\by V.V. Ivanov
\paper Geometric properties of monotone functions and probabilities of
random fluctuations.
(Translated from Sibirski\u\i\ Mat. Zh., {\bf 37}, 117--150 (1996))
\jour Siberian Mathematical Journal
\vol 37
\pages 102--129
\yr 1996
\endref
\ref
\no K
\by A.G. Kachurovskii
\paper The rate of convergence in ergodic theory.
(Translated from Uspekhi Mat. Nauk {\bf 51}, 73--124 (1996))
\jour Russian Math. Surveys
\vol 51
\pages 653--703
\yr 1996
\endref
\ref
\no RN
\by F. Riesz and B. Sz.-Nagy
\book Le\c cons d'analyse fonctionelle
\publisher Acad\'emie des Sciences de Hongrie
\yr 1955
\endref
\ref
\no R
\by D. Ruelle
\book Statistical Mechanics
\publisher New York, Addison-Wesley
\yr 1968
\endref
}

\bye

%% file: newhead.tex
\gdef\islinuxolivetti{F}
\magnification\magstep1

\newdimen\papwidth
\newdimen\papheight
\newskip\beforesectionskipamount  
\newskip\sectionskipamount 
\def\sectionskip{\vskip\sectionskipamount}
\def\beforesectionskip{\vskip\beforesectionskipamount}
\papwidth=16truecm
\papheight=22truecm
\voffset=0.4truecm
\hoffset=0.4truecm
\hsize=\papwidth
\vsize=\papheight
\nopagenumbers
\headline={\ifnum\pageno>1 {\hss\tenrm-\ \folio\ -\hss} \else
{\hfill}\fi}
\newdimen\texpscorrection
\texpscorrection=0.15truecm 

\def\sectionsize{\twelvepoint}
\def\sectiontype{\bf}
\def\subsectionsize{}
\def\subsectiontype{\bf}
\def\em{\sl}
\newfam\truecmsy
\newfam\truecmr
\newfam\msbfam
\newfam\scriptfam
\newfam\truecmsy
\newskip\ttglue 
\if T\islinuxolivetti
\papheight=11truecm

\font\twelverm=cmr12
\font\tenrm=cmr10
\font\eightrm=cmr8
\font\sevenrm=cmr7
\font\sixrm=cmr6
\font\fiverm=cmr5

\font\twelvebf=cmbx12
\font\tenbf=cmbx10
\font\eightbf=cmbx8
\font\sevenbf=cmbx7
\font\sixbf=cmbx6
\font\fivebf=cmbx5

\font\twelveit=cmti12
\font\tenit=cmti10
\font\eightit=cmti8
\font\sevenit=cmti7
\font\sixit=cmti6
\font\fiveit=cmti5

\font\twelvesl=cmsl12
\font\tensl=cmsl10
\font\eightsl=cmsl8
\font\sevensl=cmsl7
\font\sixsl=cmsl6
\font\fivesl=cmsl5

\font\twelvei=cmmi12
\font\teni=cmmi10
\font\eighti=cmmi8
\font\seveni=cmmi7
\font\sixi=cmmi6
\font\fivei=cmmi5

\font\twelvesy=cmsy10	at	12pt
\font\tensy=cmsy10
\font\eightsy=cmsy8
\font\sevensy=cmsy7
\font\sixsy=cmsy6
\font\fivesy=cmsy5
\font\twelvetruecmsy=cmsy10	at	12pt
\font\tentruecmsy=cmsy10
\font\eighttruecmsy=cmsy8
\font\seventruecmsy=cmsy7
\font\sixtruecmsy=cmsy6
\font\fivetruecmsy=cmsy5

\font\twelvetruecmr=cmr12
\font\tentruecmr=cmr10
\font\eighttruecmr=cmr8
\font\seventruecmr=cmr7
\font\sixtruecmr=cmr6
\font\fivetruecmr=cmr5

\font\twelvebf=cmbx12
\font\tenbf=cmbx10
\font\eightbf=cmbx8
\font\sevenbf=cmbx7
\font\sixbf=cmbx6
\font\fivebf=cmbx5

\font\twelvett=cmtt12
\font\tentt=cmtt10
\font\eighttt=cmtt8

\font\twelveex=cmex10	at	12pt
\font\tenex=cmex10

\font\twelvemsb=msbm10	at	12pt
\font\tenmsb=msbm10
\font\eightmsb=msbm8
\font\sevenmsb=msbm7
\font\sixmsb=msbm6
\font\fivemsb=msbm5

\font\twelvescr=eusm10 at 12pt
\font\tenscr=eusm10
\font\eightscr=eusm8
\font\sevenscr=eusm7
\font\sixscr=eusm6
\font\fivescr=eusm5
\fi
\if F\islinuxolivetti
\font\twelverm=ptmr	at	12pt
\font\tenrm=ptmr	at	10pt
\font\eightrm=ptmr	at	8pt
\font\sevenrm=ptmr	at	7pt
\font\sixrm=ptmr	at	6pt
\font\fiverm=ptmr	at	5pt

\font\twelvebf=ptmb	at	12pt
\font\tenbf=ptmb	at	10pt
\font\eightbf=ptmb	at	8pt
\font\sevenbf=ptmb	at	7pt
\font\sixbf=ptmb	at	6pt
\font\fivebf=ptmb	at	5pt

\font\twelveit=ptmri	at	12pt
\font\tenit=ptmri	at	10pt
\font\eightit=ptmri	at	8pt
\font\sevenit=ptmri	at	7pt
\font\sixit=ptmri	at	6pt
\font\fiveit=ptmri	at	5pt

\font\twelvesl=ptmro	at	12pt
\font\tensl=ptmro	at	10pt
\font\eightsl=ptmro	at	8pt
\font\sevensl=ptmro	at	7pt
\font\sixsl=ptmro	at	6pt
\font\fivesl=ptmro	at	5pt

\font\twelvei=cmmi12
\font\teni=cmmi10
\font\eighti=cmmi8
\font\seveni=cmmi7
\font\sixi=cmmi6
\font\fivei=cmmi5

\font\twelvesy=cmsy10	at	12pt
\font\tensy=cmsy10
\font\eightsy=cmsy8
\font\sevensy=cmsy7
\font\sixsy=cmsy6
\font\fivesy=cmsy5
\font\twelvetruecmsy=cmsy10	at	12pt
\font\tentruecmsy=cmsy10
\font\eighttruecmsy=cmsy8
\font\seventruecmsy=cmsy7
\font\sixtruecmsy=cmsy6
\font\fivetruecmsy=cmsy5

\font\twelvetruecmr=cmr12
\font\tentruecmr=cmr10
\font\eighttruecmr=cmr8
\font\seventruecmr=cmr7
\font\sixtruecmr=cmr6
\font\fivetruecmr=cmr5

\font\twelvebf=cmbx12
\font\tenbf=cmbx10
\font\eightbf=cmbx8
\font\sevenbf=cmbx7
\font\sixbf=cmbx6
\font\fivebf=cmbx5

\font\twelvett=cmtt12
\font\tentt=cmtt10
\font\eighttt=cmtt8

\font\twelveex=cmex10	at	12pt
\font\tenex=cmex10

\font\twelvemsb=msbm10	at	12pt
\font\tenmsb=msbm10
\font\eightmsb=msbm8
\font\sevenmsb=msbm7
\font\sixmsb=msbm6
\font\fivemsb=msbm5

\font\twelvescr=eusm10 at 12pt
\font\tenscr=eusm10
\font\eightscr=eusm8
\font\sevenscr=eusm7
\font\sixscr=eusm6
\font\fivescr=eusm5
\fi
\def\eightpoint{\def\rm{\fam0\eightrm}%
\textfont0=\eightrm
  \scriptfont0=\sixrm
  \scriptscriptfont0=\fiverm 
\textfont1=\eighti
  \scriptfont1=\sixi
  \scriptscriptfont1=\fivei 
\textfont2=\eightsy
  \scriptfont2=\sixsy
  \scriptscriptfont2=\fivesy 
\textfont3=\tenex
  \scriptfont3=\tenex
  \scriptscriptfont3=\tenex 
\textfont\itfam=\eightit
  \scriptfont\itfam=\sixit
  \scriptscriptfont\itfam=\fiveit 
  \def\it{\fam\itfam\eightit}%
\textfont\slfam=\eightsl
  \scriptfont\slfam=\sixsl
  \scriptscriptfont\slfam=\fivesl 
  \def\sl{\fam\slfam\eightsl}%
\textfont\ttfam=\eighttt
  \def\tt{\fam\ttfam\eighttt}%
\textfont\bffam=\eightbf
  \scriptfont\bffam=\sixbf
  \scriptscriptfont\bffam=\fivebf
  \def\bf{\fam\bffam\eightbf}%
\textfont\scriptfam=\eightscr
  \scriptfont\scriptfam=\sixscr
  \scriptscriptfont\scriptfam=\fivescr
  \def\script{\fam\scriptfam\eightscr}%
\textfont\msbfam=\eightmsb
  \scriptfont\msbfam=\sixmsb
  \scriptscriptfont\msbfam=\fivemsb
  \def\bb{\fam\msbfam\eightmsb}%
\textfont\truecmr=\eighttruecmr
  \scriptfont\truecmr=\sixtruecmr
  \scriptscriptfont\truecmr=\fivetruecmr
  \def\truerm{\fam\truecmr\eighttruecmr}%
\textfont\truecmsy=\eighttruecmsy
  \scriptfont\truecmsy=\sixtruecmsy
  \scriptscriptfont\truecmsy=\fivetruecmsy
\tt \ttglue=.5em plus.25em minus.15em 
\normalbaselineskip=9pt
\setbox\strutbox=\hbox{\vrule height7pt depth2pt width0pt}%
\normalbaselines
\rm
}

\def\tenpoint{\def\rm{\fam0\tenrm}%
\textfont0=\tenrm
  \scriptfont0=\sevenrm
  \scriptscriptfont0=\fiverm 
\textfont1=\teni
  \scriptfont1=\seveni
  \scriptscriptfont1=\fivei 
\textfont2=\tensy
  \scriptfont2=\sevensy
  \scriptscriptfont2=\fivesy 
\textfont3=\tenex
  \scriptfont3=\tenex
  \scriptscriptfont3=\tenex 
\textfont\itfam=\tenit
  \scriptfont\itfam=\sevenit
  \scriptscriptfont\itfam=\fiveit 
  \def\it{\fam\itfam\tenit}%
\textfont\slfam=\tensl
  \scriptfont\slfam=\sevensl
  \scriptscriptfont\slfam=\fivesl 
  \def\sl{\fam\slfam\tensl}%
\textfont\ttfam=\tentt
  \def\tt{\fam\ttfam\tentt}%
\textfont\bffam=\tenbf
  \scriptfont\bffam=\sevenbf
  \scriptscriptfont\bffam=\fivebf
  \def\bf{\fam\bffam\tenbf}%
\textfont\scriptfam=\tenscr
  \scriptfont\scriptfam=\sevenscr
  \scriptscriptfont\scriptfam=\fivescr
  \def\script{\fam\scriptfam\tenscr}%
\textfont\msbfam=\tenmsb
  \scriptfont\msbfam=\sevenmsb
  \scriptscriptfont\msbfam=\fivemsb
  \def\bb{\fam\msbfam\tenmsb}%
\textfont\truecmr=\tentruecmr
  \scriptfont\truecmr=\seventruecmr
  \scriptscriptfont\truecmr=\fivetruecmr
  \def\truerm{\fam\truecmr\tentruecmr}%
\textfont\truecmsy=\tentruecmsy
  \scriptfont\truecmsy=\seventruecmsy
  \scriptscriptfont\truecmsy=\fivetruecmsy
\tt \ttglue=.5em plus.25em minus.15em 
\normalbaselineskip=12pt
\setbox\strutbox=\hbox{\vrule height8.5pt depth3.5pt width0pt}%
\normalbaselines
\rm
}

\def\twelvepoint{\def\rm{\fam0\twelverm}%
\textfont0=\twelverm
  \scriptfont0=\tenrm
  \scriptscriptfont0=\eightrm 
\textfont1=\twelvei
  \scriptfont1=\teni
  \scriptscriptfont1=\eighti 
\textfont2=\twelvesy
  \scriptfont2=\tensy
  \scriptscriptfont2=\eightsy 
\textfont3=\twelveex
  \scriptfont3=\twelveex
  \scriptscriptfont3=\twelveex 
\textfont\itfam=\twelveit
  \scriptfont\itfam=\tenit
  \scriptscriptfont\itfam=\eightit 
  \def\it{\fam\itfam\twelveit}%
\textfont\slfam=\twelvesl
  \scriptfont\slfam=\tensl
  \scriptscriptfont\slfam=\eightsl 
  \def\sl{\fam\slfam\twelvesl}%
\textfont\ttfam=\twelvett
  \def\tt{\fam\ttfam\twelvett}%
\textfont\bffam=\twelvebf
  \scriptfont\bffam=\tenbf
  \scriptscriptfont\bffam=\eightbf
  \def\bf{\fam\bffam\twelvebf}%
\textfont\scriptfam=\twelvescr
  \scriptfont\scriptfam=\tenscr
  \scriptscriptfont\scriptfam=\eightscr
  \def\script{\fam\scriptfam\twelvescr}%
\textfont\msbfam=\twelvemsb
  \scriptfont\msbfam=\tenmsb
  \scriptscriptfont\msbfam=\eightmsb
  \def\bb{\fam\msbfam\twelvemsb}%
\textfont\truecmr=\twelvetruecmr
  \scriptfont\truecmr=\tentruecmr
  \scriptscriptfont\truecmr=\eighttruecmr
  \def\truerm{\fam\truecmr\twelvetruecmr}%
\textfont\truecmsy=\twelvetruecmsy
  \scriptfont\truecmsy=\tentruecmsy
  \scriptscriptfont\truecmsy=\eighttruecmsy
\tt \ttglue=.5em plus.25em minus.15em 
\setbox\strutbox=\hbox{\vrule height7pt depth2pt width0pt}%
\normalbaselineskip=15pt
\normalbaselines
\rm
}
%
\fontdimen16\tensy=2.7pt
\fontdimen13\tensy=4.3pt
\fontdimen17\tensy=2.7pt
\fontdimen14\tensy=4.3pt
\fontdimen18\tensy=4.3pt
\fontdimen16\eightsy=2.7pt
\fontdimen13\eightsy=4.3pt
\fontdimen17\eightsy=2.7pt
\fontdimen14\eightsy=4.3pt
\fontdimen18\eightsy=4.3pt
%
\def\hexnumber#1{\ifcase#1 0\or1\or2\or3\or4\or5\or6\or7\or8\or9\or
 A\or B\or C\or D\or E\or F\fi}
\mathcode`\=="3\hexnumber\truecmr3D
\mathchardef\not="3\hexnumber\truecmsy36
\mathcode`\+="2\hexnumber\truecmr2B
\mathcode`\(="4\hexnumber\truecmr28
\mathcode`\)="5\hexnumber\truecmr29
\mathcode`\!="5\hexnumber\truecmr21
\mathcode`\(="4\hexnumber\truecmr28
\mathcode`\)="5\hexnumber\truecmr29

\def\Phi{\mathchar"0\hexnumber\truecmr08 }
\def\Gamma {\mathchar"0\hexnumber\truecmr00 }
\def\Delta {\mathchar"0\hexnumber\truecmr01 }
\def\Theta {\mathchar"0\hexnumber\truecmr02 }
\def\Lambda{\mathchar"0\hexnumber\truecmr03 }
\def\Xi {\mathchar"0\hexnumber\truecmr04 }
\def\Pi{\mathchar"0\hexnumber\truecmr05 }
\def\Sigma{\mathchar"0\hexnumber\truecmr06 }
\def\Upsilon {\mathchar"0\hexnumber\truecmr07 }
\def\Phi {\mathchar"0\hexnumber\truecmr08 }
\def\Psi {\mathchar"0\hexnumber\truecmr09 }
\def\Omega{\mathchar"0\hexnumber\truecmr0A }
\newcount\EQNcount \EQNcount=1
\newcount\CLAIMcount \CLAIMcount=1
\newcount\SECTIONcount \SECTIONcount=0
\newcount\SUBSECTIONcount \SUBSECTIONcount=1
\def\ifff(#1,#2,#3){\ifundefined{#1#2}%
\expandafter\xdef\csname #1#2\endcsname{#3}\else%
\immediate\write16{!!!!!doubly defined #1,#2}\fi}
\def\NEWDEF #1,#2,#3 {\ifff({#1},{#2},{#3})}
\def\actualnumber{\number\SECTIONcount}
\def\EQ(#1){\lmargin(#1)\eqno\tag(#1)}
\def\NR(#1){&\lmargin(#1)\tag(#1)\cr}  
\def\tag(#1){\lmargin(#1)({\rm \actualnumber}.\number\EQNcount)
 \NEWDEF e,#1,(\actualnumber.\number\EQNcount)
\global\advance\EQNcount by 1
}
\def\SECT(#1)#2\par{\lmargin(#1)\SECTION#2\par
\NEWDEF s,#1,{\actualnumber}
}
\def\SUBSECT(#1)#2\par{\lmargin(#1)
\SUBSECTION#2\par 
\NEWDEF s,#1,{\actualnumber.\number\SUBSECTIONcount}
}
\def\CLAIM #1(#2) #3\par{
\vskip.1in\medbreak\noindent
{\lmargin(#2)\bf #1~\actualnumber.\number\CLAIMcount.} {\sl #3}\par
\NEWDEF c,#2,{#1~\actualnumber.\number\CLAIMcount}
\global\advance\CLAIMcount by 1
\ifdim\lastskip<\medskipamount
\removelastskip\penalty55\medskip\fi}
\def\CLAIMNONR #1(#2) #3\par{
\vskip.1in\medbreak\noindent
{\lmargin(#2)\bf #1.} {\sl #3}\par
\NEWDEF c,#2,{#1}
\global\advance\CLAIMcount by 1
\ifdim\lastskip<\medskipamount
\removelastskip\penalty55\medskip\fi}
\def\SECTION#1\par{\vskip0pt plus.3\vsize\penalty-75
    \vskip0pt plus -.3\vsize
    \global\advance\SECTIONcount by 1
    \beforesectionskip\noindent
{\sectionsize\sectiontype \actualnumber.\ #1}
    \EQNcount=1
    \CLAIMcount=1
    \SUBSECTIONcount=1
    \nobreak\sectionskip\noindent}
\def\SECTIONNONR#1\par{\vskip0pt plus.3\vsize\penalty-75
    \vskip0pt plus -.3\vsize
    \global\advance\SECTIONcount by 1
    \beforesectionskip\noindent
{\sectionsize\sectiontype  #1}
     \EQNcount=1
     \CLAIMcount=1
     \SUBSECTIONcount=1
     \nobreak\sectionskip\noindent}
\def\SUBSECTION#1\par{\vskip0pt plus.2\vsize\penalty-75%
    \vskip0pt plus -.2\vsize%
    \beforesectionskip\noindent%
{\subsectionsize\subsectiontype \actualnumber.\number\SUBSECTIONcount.\ #1}
    \global\advance\SUBSECTIONcount by 1
    \nobreak\sectionskip\noindent}
\def\SUBSECTIONNONR#1\par{\vskip0pt plus.2\vsize\penalty-75
    \vskip0pt plus -.2\vsize
\beforesectionskip\noindent
{\subsectionsize\subsectiontype #1}
    \nobreak\sectionskip\noindent\noindent}
\def\ifundefined#1{\expandafter\ifx\csname#1\endcsname\relax}
\def\equ(#1){\ifundefined{e#1}$\spadesuit$#1\else\csname e#1\endcsname\fi}
\def\clm(#1){\ifundefined{c#1}$\spadesuit$#1\else\csname c#1\endcsname\fi}
\def\sec(#1){\ifundefined{s#1}$\spadesuit$#1
\else Section \csname s#1\endcsname\fi}
\let\endarg=\par
\def\finish{\def\endarg{\par\endgroup}}
\def\start{\endarg\begingroup}

 \def\beginFROM{\start\parskip=0pt\vskip\baselineskip
\def\finish{\def\endarg{\egroup\par\endgroup}}
  \vbox\bgroup\obeylines\eightpoint\em\finish}

\def\ABSTRACT#1\par{
\vskip 1in {\noindent\sectionsize\sectiontype Abstract.} #1 \par}

\def\TODAY{\number\day~\ifcase\month\or January \or February \or March \or
April \or May \or June
\or July \or August \or September \or October \or November \or December \fi
\number\year\timecount=\number\time
\divide\timecount by 60
}
\newcount\timecount
\def\DRAFT{\def\lmargin(##1){\strut\vadjust{\kern-\strutdepth
\vtop to \strutdepth{
\baselineskip\strutdepth\vss\rlap{\kern-1.2 truecm\eightpoint{##1}}}}}
\font\footfont=cmti7
\footline={{\footfont \hfil File:\jobname, \TODAY,  \number\timecount h}}
}
\newbox\strutboxJPE
\setbox\strutboxJPE=\hbox{\strut}
\def\subitem#1#2\par{\vskip\baselineskip\vskip-\ht\strutboxJPE{\item{#1}#2}}
\gdef\strutdepth{\dp\strutbox}
\def\lmargin(#1){}
\def\period{\unskip.\spacefactor3000 { }}
%
%
\newbox\noboxJPE
\newbox\byboxJPE
\newbox\paperboxJPE
\newbox\yrboxJPE
\newbox\jourboxJPE
\newbox\pagesboxJPE
\newbox\volboxJPE
\newbox\preprintboxJPE
\newbox\toappearboxJPE
\newbox\bookboxJPE
\newbox\bybookboxJPE
\newbox\publisherboxJPE
\newbox\inprintboxJPE
\def\refclearJPE{
   \setbox\noboxJPE=\null             \gdef\isnoJPE{F}
   \setbox\byboxJPE=\null             \gdef\isbyJPE{F}
   \setbox\paperboxJPE=\null          \gdef\ispaperJPE{F}
   \setbox\yrboxJPE=\null             \gdef\isyrJPE{F}
   \setbox\jourboxJPE=\null           \gdef\isjourJPE{F}
   \setbox\pagesboxJPE=\null          \gdef\ispagesJPE{F}
   \setbox\volboxJPE=\null            \gdef\isvolJPE{F}
   \setbox\preprintboxJPE=\null       \gdef\ispreprintJPE{F}
   \setbox\toappearboxJPE=\null       \gdef\istoappearJPE{F}
   \setbox\inprintboxJPE=\null        \gdef\isinprintJPE{F}
   \setbox\bookboxJPE=\null           \gdef\isbookJPE{F}  \gdef\isinbookJPE{F}
     
   \setbox\bybookboxJPE=\null         \gdef\isbybookJPE{F}
   \setbox\publisherboxJPE=\null      \gdef\ispublisherJPE{F}
     
}
\def\widestlabel#1{\setbox0=\hbox{#1\enspace}\refindent=\wd0\relax}
\def\ref{\refclearJPE\bgroup}
\def\no   {\egroup\gdef\isnoJPE{T}\setbox\noboxJPE=\hbox\bgroup}
\def\by   {\egroup\gdef\isbyJPE{T}\setbox\byboxJPE=\hbox\bgroup}
\def\paper{\egroup\gdef\ispaperJPE{T}\setbox\paperboxJPE=\hbox\bgroup}
\def\yr{\egroup\gdef\isyrJPE{T}\setbox\yrboxJPE=\hbox\bgroup}
\def\jour{\egroup\gdef\isjourJPE{T}\setbox\jourboxJPE=\hbox\bgroup}
\def\pages{\egroup\gdef\ispagesJPE{T}\setbox\pagesboxJPE=\hbox\bgroup}
\def\vol{\egroup\gdef\isvolJPE{T}\setbox\volboxJPE=\hbox\bgroup\bf}
\def\preprint{\egroup\gdef
\ispreprintJPE{T}\setbox\preprintboxJPE=\hbox\bgroup}
\def\toappear{\egroup\gdef
\istoappearJPE{T}\setbox\toappearboxJPE=\hbox\bgroup}
\def\inprint{\egroup\gdef
\isinprintJPE{T}\setbox\inprintboxJPE=\hbox\bgroup}
\def\book{\egroup\gdef\isbookJPE{T}\setbox\bookboxJPE=\hbox\bgroup\em}
\def\publisher{\egroup\gdef
\ispublisherJPE{T}\setbox\publisherboxJPE=\hbox\bgroup}
\def\inbook{\egroup\gdef\isinbookJPE{T}\setbox\bookboxJPE=\hbox\bgroup\em}
\def\bybook{\egroup\gdef\isbybookJPE{T}\setbox\bybookboxJPE=\hbox\bgroup}
\newdimen\refindent
\refindent=5em
\def\endref{\egroup \sfcode`.=1000
 \if T\isnoJPE
 \hangindent\refindent\hangafter=1
      \noindent\hbox to\refindent{[\unhbox\noboxJPE\unskip]\hss}\ignorespaces
     \else  \noindent    \fi
 \if T\isbyJPE    \unhbox\byboxJPE\unskip: \fi
 \if T\ispaperJPE \unhbox\paperboxJPE\unskip\period \fi
 \if T\isbookJPE {\it\unhbox\bookboxJPE\unskip}\if T\ispublisherJPE, \else.
\fi\fi
 \if T\isinbookJPE In {\it\unhbox\bookboxJPE\unskip}\if T\isbybookJPE,
\else\period \fi\fi
 \if T\isbybookJPE  (\unhbox\bybookboxJPE\unskip)\period \fi
 \if T\ispublisherJPE \unhbox\publisherboxJPE\unskip \if T\isjourJPE, \else\if
T\isyrJPE \  \else\period \fi\fi\fi
 \if T\istoappearJPE (To appear)\period \fi
 \if T\ispreprintJPE Pre\-print\period \fi
 \if T\isjourJPE    \unhbox\jourboxJPE\unskip\ \fi
 \if T\isvolJPE     \unhbox\volboxJPE\unskip\if T\ispagesJPE, \else\ \fi\fi
 \if T\ispagesJPE   \unhbox\pagesboxJPE\unskip\  \fi
 \if T\isyrJPE      (\unhbox\yrboxJPE\unskip)\period \fi
 \if T\isinprintJPE (in print)\period \fi
\filbreak
}
\def\hexnumber#1{\ifcase#1 0\or1\or2\or3\or4\or5\or6\or7\or8\or9\or
 A\or B\or C\or D\or E\or F\fi}
\textfont\msbfam=\tenmsb
\scriptfont\msbfam=\sevenmsb
\scriptscriptfont\msbfam=\fivemsb
\mathchardef\varkappa="0\hexnumber\msbfam7B
\newcount\FIGUREcount \FIGUREcount=0
\newdimen\figcenter
\def\fig(#1){\ifundefined{fig#1}%
\global\advance\FIGUREcount by 1%
\NEWDEF fig,#1,{Fig.~\number\FIGUREcount}%
\immediate\write16{ FIG \number\FIGUREcount : #1}
\fi
\csname fig#1\endcsname\relax}
\def\figure #1 #2 #3 #4\cr{\null%
\ifundefined{fig#1}%
\global\advance\FIGUREcount by 1%
\NEWDEF fig,#1,{\number\FIGUREcount}
\immediate\write16{  FIG \number\FIGUREcount : #1}
\fi
{\goodbreak\figcenter=\hsize\relax
\advance\figcenter by -#3truecm
\divide\figcenter by 2
\midinsert\vskip #2truecm\noindent\hskip\figcenter
\includegraphics{#1}\vskip 0.8truecm\noindent \vbox{\eightpoint\noindent
{\bf\fig(#1)}: #4}\endinsert}}
\def\figurewithtex #1 #2 #3 #4 #5\cr{\null%
\ifundefined{fig#1}%
\global\advance\FIGUREcount by 1%
\NEWDEF fig,#1,{\csname fig#1\endcsname}
\immediate\write16{ FIG \number\FIGUREcount: #1}
\fi
{\goodbreak\figcenter=\hsize\relax
\advance\figcenter by -#4truecm
\divide\figcenter by 2
\midinsert\vskip #3truecm\noindent\hskip\figcenter
\includegraphics{#1}{\hskip\texpscorrection\input #2 }\vskip 0.8truecm\noindent \vbox{\eightpoint\noindent
{\bf\fig(#1)}: #5}\endinsert}}
\def\figurewithtexplus #1 #2 #3 #4 #5 #6\cr{\null%
\ifundefined{fig#1}%
\global\advance\FIGUREcount by 1%
\NEWDEF fig,#1,{Fig.~\number\FIGUREcount}
\immediate\write16{ FIG \number\FIGUREcount: #1}
\fi
{\goodbreak\figcenter=\hsize\relax
\advance\figcenter by -#4truecm
\divide\figcenter by 2
\midinsert\vskip #3truecm\noindent\hskip\figcenter
\includegraphics{#1}{\hskip\texpscorrection\input #2 }\vskip #5truecm\noindent \vbox{\eightpoint\noindent
{\bf\fig(#1)}: #6}\endinsert}}
\catcode`@=11
\def\footnote#1{\let\@sf\empty 
  \ifhmode\edef\@sf{\spacefactor\the\spacefactor}\/\fi
  #1\@sf\vfootnote{#1}}
\def\vfootnote#1{\insert\footins\bgroup\eightpoint
  \interlinepenalty\interfootnotelinepenalty
  \splittopskip\ht\strutbox 
  \splitmaxdepth\dp\strutbox \floatingpenalty\@MM
  \leftskip\z@skip \rightskip\z@skip \spaceskip\z@skip \xspaceskip\z@skip
  \textindent{#1}\footstrut\futurelet\next\fo@t}
\def\fo@t{\ifcat\bgroup\noexpand\next \let\next\f@@t
  \else\let\next\f@t\fi \next}
\def\f@@t{\bgroup\aftergroup\@foot\let\next}
\def\f@t#1{#1\@foot}
\def\@foot{\strut\egroup}
\def\footstrut{\vbox to\splittopskip{}}
\skip\footins=\bigskipamount 
\count\footins=1000 
\dimen\footins=8in 
\catcode`@=12 

\def\CC{{\script C}}

\def\OO{{\script O}}


\def\QED{\hfill\smallskip
         \line{$\hfill{\vcenter{\vbox{\hrule height 0.2pt
	\hbox{\vrule width 0.2pt height 1.8ex \kern 1.8ex
		\vrule width 0.2pt}
	\hrule height 0.2pt}}}$
               \ \ \ \ \ \ }
         \bigskip}
\def\real{{\bf R}}
\def\natural{{\bf N}}

\def\integer{{\bf Z}}

\def\PROOF{\medskip\noindent{\bf Proof.\ }}
\def\REMARK{\medskip\noindent{\bf Remark.\ }}
\def\LIKEREMARK#1{\medskip\noindent{\bf #1.\ }}
\tenpoint
\normalbaselineskip=5.25mm
\baselineskip=5.25mm
\parskip=10pt
\beforesectionskipamount=24pt plus8pt minus8pt
\sectionskipamount=3pt plus1pt minus1pt
\def\em{\it}

%% file: layout.tex
\normalbaselineskip=12pt
\baselineskip=12pt
\parskip=0pt
\parindent=22.222pt
\beforesectionskipamount=24pt plus0pt minus6pt
\sectionskipamount=7pt plus3pt minus0pt
\overfullrule=0pt
\hfuzz=2pt
\nopagenumbers
\headline={\ifnum\pageno>1 {\hss\tenrm-\ \folio\ -\hss} \else
{\hfill}\fi}
\if T\islinuxolivetti
\font\titlefont=cmbx14

\font\toplinefont=cmcsc10
\font\pagenumberfont=cmbx10
\else
\font\titlefont=ptmb at 18pt

\font\toplinefont=cmcsc10
\font\pagenumberfont=ptmb at 10pt
\fi
\newdimen\itemindent\itemindent=1.5em

\def\textindent#1{\indent\llap{#1\enspace}\ignorespaces}
\def\item{\par\noindent
\hangindent\itemindent\hangafter=1\relax
\setitemmark}
\def\setitemindent#1{\setbox0=\hbox{\ignorespaces#1\unskip\enspace}%
\itemindent=\wd0\relax
\message{|\string\setitemindent: Mark width modified to hold
         |`\string#1' plus an \string\enspace\space gap. }%
}
\def\setitemmark#1{\checkitemmark{#1}%
\hbox to\itemindent{\hss#1\enspace}\ignorespaces}
\def\checkitemmark#1{\setbox0=\hbox{\enspace#1}%
\ifdim\wd0>\itemindent
   \message{|\string\item: Your mark `\string#1' is too wide. }%
\fi}
\setitemindent{3.)}
\def\SECTION#1\par{\vskip0pt plus.2\vsize\penalty-75
    \vskip0pt plus -.2\vsize
    \global\advance\SECTIONcount by 1
    \beforesectionskip\noindent
{\sectionsize\sectiontype \actualnumber.\ #1}
    \EQNcount=1
    \CLAIMcount=1
    \SUBSECTIONcount=1
    \nobreak\sectionskip\noindent}